# Detoxification of Superoxide without Production of $H_2O_2$. A New Antioxidant Activity of Superoxide Reductase Complexed with Ferrocyanide

Classification: Biological Sciences, Biochemistry

Fernando P. Molina-Heredia[‡$], Chantal Houée-Levin[§], Catherine Berthomieu[†], Danièle Touati[¶], Emilie Tremey[‡], Vincent Favaudon[¶¶], Virgile Adam[#], and Vincent Nivière[‡]

[‡] DRDC/CB, UMR 5047, CEA/CNRS/Univ. J. Fourier, 17 Av. des Martyrs, 38054 Grenoble Cedex 9, France. [§] LCP, UMR 8000, CNRS/Univ. Paris-Sud, Bât. 350, 91405 Orsay Cedex, France. [†] DEVM/LIPM, UMR 6191, CEA-Cadarache, 13108 Saint Paul-lez-Durance, France. [¶] Inst. J. Monod, CNRS/Univ. Paris 6 et 7, 2 place Jussieu, 75251 Paris Cedex 5, France. [$] Present address: IBVF, Univ. de Sevilla y CSIC, Américo Ves. 49, 41092 Sevilla, Spain. [¶¶] Inserm U 612 and Institut Curie, Bât. 110-112, 91405 Orsay Cedex, France. [#] ESRF, BP 220, 38043 Grenoble Cedex, France.

Corresponding author: Vincent Nivière, E-mail: vniviere@cea.fr, Tel.: 33 4 38 78 91 09, Fax: 33 4 38 78 91 24.

Manuscript informations: 18 text pages

Word and character counts: 215 words in the abstract and 35 672 characters in total

[1]Abbreviations footnote: SOD, superoxide dismutase; SOR, superoxide reductase

ABSTRACT

The superoxide radical $O_2^{\bullet-}$ is a toxic by-product of oxygen metabolism. Two $O_2^{\bullet-}$ detoxifying enzymes have been described so far, superoxide dismutase and superoxide reductase, both forming $H_2O_2$ as reaction product. Recently, the superoxide reductase active site, a ferrous iron in a [$Fe^{2+}$ (N-His)$_4$ (S-Cys)] pentacoordination, was shown to have the ability to form a complex with the organometallic compound ferrocyanide. Here, we have investigated in detail the reactivity of the superoxide reductase-ferrocyanide complex with $O_2^{\bullet-}$ by pulse and γ-ray radiolysis, infrared and UV-visible spectroscopies. The complex reacts very efficiently with $O_2^{\bullet-}$. However, the presence of the ferrocyanide adduct markedly modifies the reaction mechanism of superoxide reductase, with the formation of transient intermediates different from those observed for SOR alone. A one-electron redox chemistry appears to be carried out by the ferrocyanide moiety of the complex, while the superoxide reductase iron site remains in the reduced state. Surprisingly, the toxic $H_2O_2$ species is no longer the reaction product. Accordingly, *in vivo* experiments showed that formation of the superoxide reductase ferrocyanide complex increased the antioxidant capabilities of superoxide reductase expressed in an *Escherichia coli sodA sodB recA* mutant strain. Altogether, these data describe an unprecedented $O_2^{\bullet-}$ detoxification activity, catalyzed by the superoxide reductase-ferrocyanide complex, which does not conduct to the production of the toxic $H_2O_2$ species.

INTRODUCTION

The superoxide radical $O_2^{\bullet-}$, the one-electron reduction product of molecular oxygen, is a toxic by-product of oxygen metabolism (1, 2). Fortunately, cells possess $O_2^{\bullet-}$ detoxifying enzymes, which are essential to their survival in the presence of oxygen. Two different types of $O_2^{\bullet-}$ detoxifying enzymes have been described so far. The first one is the well-known metalloenzyme superoxide dismutase (SOD)[1], present in almost all aerobic cells (1, 2). It catalyzes the dismutation of $O_2^{\bullet-}$ into $H_2O_2$ and $O_2$:

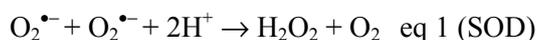

$O_2^{\bullet-} + O_2^{\bullet-} + 2H^+ \rightarrow H_2O_2 + O_2$   eq 1 (SOD)

The second system, only found in prokaryotic cells, is the recently characterized non-heme iron superoxide reductase (SOR), that catalyzes reduction of $O_2^{\bullet-}$ into $H_2O_2$ (3-6):

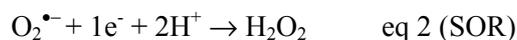
$$O_2^{\bullet-} + 1e^- + 2H^+ \rightarrow H_2O_2 \qquad \text{eq 2 (SOR)}$$

Although both systems produce $H_2O_2$, a toxic substance precursor of $HO^{\bullet}$ radicals that needs to be eliminated by catalase or peroxidase enzymes, the elimination of $O_2^{\bullet-}$ by SOD or SOR appears to be an essential cellular process to allow organisms to survive in the presence of $O_2$ (1, 2, 6).

The SOR active site consists of a mononuclear ferrous ion in an unusual [$Fe^{2+}$ (N-His)$_4$ (S-Cys)] square pyramidal pentacoordination (7, 8). The free, solvent exposed, sixth coordination position is the site of $O_2^{\bullet-}$ reduction (5, 6). In the case of the SOR from *Desulfoarculus baarsii*, the reaction with $O_2^{\bullet-}$ proceeds through two reaction intermediates (9, 10). The first one, presumably a $Fe^{3+}$-peroxo species, is formed by the almost diffusion limited binding of $O_2^{\bullet-}$ to the ferrous active site ($k=1 \times 10^9$ $M^{-1}s^{-1}$). This intermediate undergoes two sequential protonation processes, first yielding a second intermediate, possibly a $Fe^{3+}$-hydroperoxo species and then the final reaction products, $H_2O_2$ and the ferric active site (10).

Recently, spectroscopic (11, 12) and crystallographic studies (8) have shown that the SOR active site binds ferrocyanide (also referred as hexacyanoferrate (II) or $K_4Fe(CN)_6$) at its sixth coordination position through a cyano-bridge between the iron and the ferrocyanide molecule (Fig. 1). The complex was observed for both the reduced and oxidized forms of the SOR iron active site (8). The almost perfect steric and electrostatic complementarities of ferrocyanide with the active site suggested that it could act as a potent inhibitor of SOR activity, by preventing $O_2^{\bullet-}$ access to the iron.

This SOR-ferrocyanide adduct is the first complex between an organometallic compound and the iron site of a metalloenzyme that has been described to date. Here, we have studied in detail the reaction mechanism of this complex with $O_2^{\bullet-}$. We found that the complex reacts efficiently with $O_2^{\bullet-}$, in a different way than SOR alone, and eliminates $O_2^{\bullet-}$ without production of toxic $H_2O_2$. Accordingly, formation of the SOR-ferrocyanide complex in vivo specifically increases the protective effect of SOR expression in *Escherichia coli*. This is the first report of such an antioxidant activity.

RESULTS

*Reaction mechanism of the SOR-ferrocyanide complex studied by pulse radiolysis*

The reaction of *D. baarsii* SOR with $O_2^{\bullet-}$ in the presence of potassium ferrocyanide was investigated by pulse radiolysis, as previously described (9, 10). $O_2^{\bullet-}$ was generated in less than 1 μs by scavenging radiolytically generated $HO^{\bullet}$ free radicals by 100 mM formate, in $O_2$ saturated solution, at pH 7.6. Aliquots of concentrated ferrocyanide were added to the solution just prior to the irradiation. The SOR protein was present in large excess with regard to $O_2^{\bullet-}$ ([SOR]=100 μM, $[O_2^{\bullet-}]$=3 μM), to provide pseudo first-order conditions. The kinetics of the reaction of SOR with $O_2^{\bullet-}$ were investigated every 5 to 10 nm between 395 and 660 nm, in the presence of 0, 0.25, 0.5, 1, 1.2, 1.5, 2, 5, 7, and 10 molar equivalents of ferrocyanide with respect to SOR.

Fig. 2A shows representative kinetic traces of the reaction in the 0.1 ms time scale, recorded at 580 nm in the presence of 0, 1, and 1.2 to 10 molar equivalent of ferrocyanide. In the absence of ferrocyanide, as described previously (9, 10), absorbances reached a maximum ca. 50 μs after the pulse to form the first reaction intermediate. At low ferrocyanide:SOR ratio, increasing the concentration of ferrocyanide slowed down the reaction and induced a decrease of the maximal absorbance variation (Fig. 2A). No further change occurred above a ferrocyanide:SOR ratio of 1.2, up to 10 molar equivalents of ferrocyanide, at all the wavelengths investigated. For a given ferrocyanide concentration, all the kinetic traces fitted a single exponential time-dependent equation, with identical rate constants $k_{1app}$. At ferrocyanide:SOR ratio <1.2 (Fig. 2B), $k_{1app}$ decreased linearly with increasing ferrocyanide concentration. Above a slight molar excess of ferrocyanide, the reaction became ferrocyanide independent, with a $k_{1app}$ value of $(1.70\pm0.08)\times10^4$ s$^{-1}$, a value about 6 times lower than in the absence of ferrocyanide (Fig. 2B). In the presence of 5 molar equivalents of ferrocyanide with respect to SOR, the $k_{1app}$ value was found to be proportional to the SOR concentration (Inset Fig. 2B). A $K_d$ value of 0.80±0.07 μM was determined for the binding of ferrocyanide to SOR (Fig. S1). These data indicate that as soon as enough ferrocyanide is present in solution to form a stoechiometric complex with SOR (ferrocyanide:SOR ratio ≥1.2), a bimolecular reaction between the SOR-ferrocyanide complex and $O_2^{\bullet-}$ occurs, with a rate constant $k_1$ value of $(1.8\pm0.1)\times10^8$ $M^{-1}s^{-1}$.

In the absence of ferrocyanide, as described previously (9, 10), a second reaction intermediate was formed ca. 10 ms after the pulse. Fig. 3 shows representative traces at 580 nm, in the presence of 0, 1, and 1.2 to 10 molar equivalents of ferrocyanide, in the millisecond timescale. Between 0 and 1.2 molar equivalent of ferrocyanide, increasing the concentration of ferrocyanide induced variation of the amplitude of the traces at the different wavelengths investigated (Fig. 3 and not shown). No further change occurred above a ferrocyanide:SOR ratio of 1.2, up to 10 molar equivalents of ferrocyanide. For ferrocyanide:SOR ratios between 1.2 and 10, after the first rapid process described by $k_1$, all the kinetic traces were single exponential processes, with identical rate constants $k_{2app}$=445±20 $s^{-1}$ (Fig. 3). In the presence of 5 equivalents of ferrocyanide with respect to SOR, this rate constant did not depend on the SOR concentration (inset of Fig. 3). This shows that the free ferrocyanide molecules present in the solution do not react with the complex at this stage. At constant ionic strength, in the presence of 5 equivalents of ferrocyanide, $k_{2app}$ value was found to vary linearly with Tris and formate concentrations (Fig. S2), whereas $k_1$ value did not depend on these concentrations (data not shown). When Tris concentration was varied between 1 and 10 mM, $k_{2app}$ increased by a factor of 1.6. More significantly, when formate concentration was varied between 10 and 100 mM, $k_{2app}$ increased by a factor of 3.5 (Fig. S2). This shows that formate, and more moderately Tris, are involved in the reaction process described by $k_{2app}$.

The data are consistent with the formation of two reaction intermediates. Their spectra were constructed at the delay time of their maximum formation. Fig. 4A shows the spectrum of the first intermediate recorded 300 μs after the pulse for a SOR:ferrocyanide ratio of 5. It differs from that recorded in the absence of ferrocyanide (10). Two peaks are observed at 420 and 580 nm. The 580 nm peak is 2.2 times lower than that of the first reaction intermediate observed without ferrocyanide at the same wavelength. The absorbance band centered at 420 nm is very similar to that of a ferricyanide ($Fe^{3+}(CN)_6$) species.

The reconstituted spectrum of the second reaction intermediate at its maximum, 10 ms after the pulse (SOR:ferrocyanide ratio of 5, Fig. 4B), also strongly differs from the corresponding one, recorded without ferrocyanide (10). Actually, it is identical to that of a ferricyanide ($Fe^{3+}(CN)_6$) species, with a single peak at 420 nm and at the same concentration as that of $O_2^{\bullet-}$ that reacted with the complex (3 µM) (Fig. 4B, dotted line).

Control experiments indicated that ferrocyanide alone (500 µM), in the absence of SOR, did not react with 3 µM of $O_2^{\bullet-}$ (data not shown). Also, ferrocyanide alone (6 µM) or in the presence of SOR (1 µM) did not catalyze dismutation of 10 µM of $O_2^{\bullet-}$, at pH 7.6 (data not shown).

*Final products of the reaction*

Continuous γ-ray irradiation experiments, in which $O_2^{\bullet-}$ is at a steady state (13), are well adapted to analyze the final products of the reaction between the SOR-ferrocyanide complex and $O_2^{\bullet-}$ (10). Irradiation of aqueous solutions always leads to $H_2O_2$ production during the primary steps of water radiolysis (the so-called radiolytic $H_2O_2$, G≈0.07 µmol $J^{-1}$ (13)). Here, the dose was equal to 88 Gy, hence the radiolytic $H_2O_2$ production was equal to 6.1 µM. As illustrated in the introduction from eqs 1 and 2, the measurement of the $H_2O_2$ production allows distinction between a reduction or a dismutation of $O_2^{\bullet-}$. In the presence of SOR alone, 55 µM of $O_2^{\bullet-}$ (yield≈0.62 µmol $J^{-1}$) were stoechiometrically reduced into $H_2O_2$ (Table 1), as previously reported (10). In the presence of increasing amounts of ferrocyanide, the $H_2O_2$ production decreased, and surprisingly, in the presence of more than one equivalent of ferrocyanide, only radiolytic $H_2O_2$ was detected. This suggests that scavenging of $O_2^{\bullet-}$ by the SOR-ferrocyanide complex does not produce $H_2O_2$.

In order to verify that the SOR-ferrocyanide solution does not lead to catalase activity, we incubated a solution containing 100 µM SOR and 5 molar equivalents of ferrocyanide with 50 µM of $H_2O_2$ for 15 min, room temperature at pH 7.6. No decrease in $H_2O_2$ concentration was observed. Also, a solution containing 100 µM SOR with 5 molar equivalents of ferrocyanide was allowed to react with 55 µM $O_2^{\bullet-}$ in the presence of 50 µM $H_2O_2$ (Table 1). At the end of the reaction, 48 µM $H_2O_2$ were detected in the reaction mixture (corrected from the radiolytic yield for $H_2O_2$). This confirms that no catalase activity was present in the solution during the reaction with $O_2^{\bullet-}$. Inhibition of the horseradish peroxidase enzyme used to quantify $H_2O_2$ was excluded, since it is still possible to detect $H_2O_2$ added before the reaction or resulting from the radiolysis. Finally, ferrocyanide alone does not trap $H_2O_2$, since $O_2^{\bullet-}$ produced during the continuous γ-ray irradiation is quantitatively dismutated into $H_2O_2$ in the presence of 200 μM of ferrocyanide, and ferrocyanide was not oxidized (Table 1).

The final difference spectra of the SOR-ferrocyanide solutions after reaction with $O_2^{\bullet-}$ (reference is unreacted solutions) allow quantification of oxidized SOR and ferricyanide (Table 1 and Fig. S3). Identical spectra were obtained in the presence of catalytic amounts of catalase (not shown). In the absence of ferrocyanide, as previously reported (10), SOR is stoechiometrically oxidized by $O_2^{\bullet-}$ (Table 1). In the presence of increasing equivalents of ferrocyanide, the concentration of oxidized SOR is lowered, whereas that of ferricyanide increases, as determined from the apparition of its absorbance band at 420 nm (Fig. S3 and Table 1). The sum of the oxidized amounts of SOR plus ferricyanide remains constant and equal to the concentration of $O_2^{\bullet-}$ produced during the experiment (Table 1). Treating these irradiated solutions with iridium chloride or ammonium persulfate induces full oxidation of SOR and ferrocyanide, the concentrations of which are identical to those before irradiation (data not shown). This indicates that neither the active site of SOR nor the $Fe(CN)_6$ species have been destroyed during the reaction.

*FTIR spectroscopy analysis of the reaction*

Fig. 5 shows FTIR difference spectra of the SOR-ferrocyanide solutions (with 2 and 5 ferrocyanide equivalents) after *minus* prior to the reaction with $O_2^{\bullet-}$. The experimental conditions were identical to those reported in Table 1. The presence of a negative band at 2037 cm$^{-1}$ and of a positive band at 2116 cm$^{-1}$ indicates that during the reaction with $O_2^{\bullet-}$, there was a net consumption of ferrocyanide together with a net formation of ferricyanide, respectively. The frequency of the IR bands are typical for free species in solution, i.e. not complexed to SOR. Quantification of ferrocyanide and ferricyanide from the 2037 and 2116 cm$^{-1}$ bands indicates that ≈18 and 20 µM of free ferrocyanide have been consumed and 15-18 and 20-23 µM of free ferricyanide have been formed, during the experiments containing 2 and 5 equivalents of ferrocyanide, respectively. The ferrocyanide consumption is nearly identical to that of the ferricyanide production, which fits well with the values reported in Table 1.

It has been reported that the SOR-ferrocyanide complex for the *Treponema pallidum* and *Desulfovibrio vulgaris* enzymes exhibits characteristic FTIR bands at 2095, 2047 and 2024 cm$^{-1}$, assigned to the ν(CN) stretching mode of the bridging, equatorial and axial cyanide groups of ferrocyanide, respectively (12). The FTIR spectrum of a concentrated preparation of a *D. baarsii* SOR-ferrocyanide complex shows almost identical features (Fig. S4). The FTIR spectrum of the SOR-ferrocyanide complex after irradiation, in the same experimental conditions as those of Table 1 (followed by concentration), shows no significant modifications of the frequency and intensities of the bands associated with the ferrocyanide complexed with SOR (Fig. S4). This shows that the SOR-ferrocyanide complex has not been altered during the reaction with $O_2^{\bullet-}$ and that the ferrocyanide bound to SOR remains in the reduced state after the reaction with $O_2^{\bullet-}$.

*Search for Fenton reaction*

It is well known that $H_2O_2$ leads to HO$^{\bullet}$ radicals in the presence of iron complexes (1, 2). If this very powerful oxidant would form during the reaction of the SOR-ferrocyanide complex with $O_2^{\bullet-}$, it would react instantaneously with the protein or the ferrocyanide present in solution. The electrospray mass spectrum of the irradiated SOR-ferrocyanide complex (5 equivalents of ferrocyanide, same experimental conditions as those of Table 1) shows that the mass of the SOR polypeptide is not modified, with only ions detected at 14,028±2 Da (data not shown). This demonstrates the absence of oxidative damage on the SOR polypeptide. Together with the UV-visible and FTIR data, it confirms that the SOR-ferrocyanide complex is not altered after the reaction with $O_2^{\bullet-}$.

*In vivo, the SOR-ferrocyanide complex increases protection against oxidative stress*

The results show that the SOR-ferrocyanide complex scavenges $O_2^{\bullet-}$ without generating $H_2O_2$. Given the high toxicity of $H_2O_2$, such a property could result in a significant increase of the antioxidant capability of SOR *in vivo*, resulting in enhanced protection of the cell against oxidative damage. We tested this hypothesis by analyzing the phenotype of an *E. coli sodA sodB recA* mutant strain in which SOR was expressed in the presence or in the absence of ferrocyanide in the culture medium (Table 2). The *E. coli* mutant did not grow in the presence of oxygen because of the combined lack of superoxide dismutase (*sodA sodB*) and the DNA strand break repair activity (*recA*), which results in lethal oxidative damage (14). Previous studies have shown that full overexpression of SOR efficiently compensates for the lack of SOD in *E. coli* (14). As shown in Table 2, moderate expression of SOR, achieved by induction of the *sor* gene with low IPTG concentrations, efficiently but not totally compensates the lack of SOD. Expression of an *E. coli* Mn SOD in the same experimental conditions leads to a similar phenotype (Table 2). When 1 mM ferrocyanide is added to the culture medium, the viability of the SOR expressing bacteria is increased by a factor of three. The difference is also significant with 5 mM ferrocyanide, which is the upper limit of apparent toxicity of ferrocyanide for *E. coli* cells (data not shown). Addition of ferrocyanide at 1 or 5 mM in the culture of the Mn SOD expressing bacteria, did not induce any significant effect on the aerobic survival of the cell (Table 2). These data show a specific effect of the presence of ferrocyanide in increasing the antioxidant properties of the SOR-expressing system *in vivo*.

Soluble extracts were prepared from cells overexpressing SOR and grown in the presence of 1 mM ferrocyanide. After oxidation using ammonium persulfate, the UV-visible analyses of the soluble extracts reveal the presence of a 420 nm band, characteristic of ferricyanide. This band is not present in extracts from cells grown in the absence of ferrocyanide (Fig. S5). In addition, size exclusion chromatography of the soluble extracts showed that ferrocyanide is associated with the SOR protein (Fig. S5). These data demonstrate that a SOR-Fe(CN)$_6$ complex is formed within the cells when ferrocyanide is added to the culture medium.

In the presence of NADH or NADPH as electron donors, cells extracts were found to catalyze the reduction of the oxidized forms of the SOR-Fe(CN)$_6$ complex (Table S1). This shows that the soluble extracts contain NAD(P)H dependent reductases that can regenerate the active form of the complex. The reduction rates of the complex were found similar to that reported for SOR alone (4). These data indicate that the reaction of the SOR-Fe(CN)$_6$ complex with $O_2^{\bullet-}$ can be catalytic within the cell.

DISCUSSION

The recent crystal structure of SOR from *D. baarsii* revealed that the organometallic compound ferrocyanide makes a complex with the enzyme active site (8). This is the first example of a complex between ferrocyanide and an iron site of a metalloenzyme. The ferrocyanide entirely plugs the SOR active site, with several surface residues involved in hydrogen bonds with the cyanide moieties (Fig. 1). This intriguing structure initially suggested that ferrocyanide could potentially be a strong inhibitor of SOR activity.

Here, we show that, on the contrary, the SOR-ferrocyanide complex reacts very efficiently with $O_2^{\bullet-}$. However, the presence of the ferrocyanide adduct markedly modifies the reaction mechanism. Although the complex, like SOR alone, carries out one-electron reduction chemistry on $O_2^{\bullet-}$, $H_2O_2$ is no longer the final reaction product. The absence of $H_2O_2$ production in the reaction of the SOR-ferrocyanide complex with $O_2^{\bullet-}$ is in full agreement with the specific increase of the antioxidant capabilities of SOR expression in an *E. coli sodA sodB recA* mutant strain when complexed with ferrocyanide.

In Scheme 1, we propose a reaction mechanism of the SOR-ferrocyanide complex with $O_2^{\bullet-}$, in which the reduction process is essentially carried out by the ferrocyanide adduct. This is supported by the following points.

Pulse radiolysis and binding experiments show that as soon as enough ferrocyanide is present in solution to saturate the SOR active site, the SOR-ferrocyanide complex reacts with $O_2^{\bullet-}$ with a bimolecular rate constant of $1.8 \times 10^8$ $M^{-1}s^{-1}$. This rate constant is about 6 times lower than that obtained in the absence of ferrocyanide, but still represents a very efficient scavenging activity towards $O_2^{\bullet-}$. From this bimolecular reaction, two intermediate species are formed sequentially, with different absorption spectra from those observed in the absence of ferrocyanide (Fig. 4, Scheme 1). In both intermediates, the presence of an absorption band centered at 420 nm shows that the ferrocyanide ($Fe^{2+}$) moiety of the complex has been oxidized into ferricyanide ($Fe^{3+}$). The additional band centered at 580 nm in the first intermediate (Fig. 4A) does not support an oxidation of the SOR iron active site, since in that case a broader band centered at 650 nm with a higher intensity would be expected (4, 10). It could be consistent with the presence of a peroxide species, as discussed below. For the second intermediate, quantification of the ferricyanide formation corresponds exactly to the amount of $O_2^{\bullet-}$ that reacted with the complex. No other absorbance bands are observed, suggesting that this intermediate corresponds to a SOR($Fe^{2+}$)-ferricyanide($Fe^{3+}$) complex. A one-electron redox chemistry is therefore carried out by the ferrocyanide moiety of the complex, while the SOR iron site remains in the reduced state (Scheme 1).

It should be noted that, because ferrocyanide is negatively charged, one would have expected strong electrostatic repulsions with the superoxide anion. However, when complexed to SOR, the electrostatic properties of the complex with the presence of positive charges near the ferrocyanide, Lys48 for example (Fig.1), could be one of the features that allow reaction of $O_2^{\bullet-}$ with the ferrocyanide adduct.

FTIR analysis of the final reaction products a few minutes after the reaction with $O_2^{\bullet-}$ show that the protein-bound $Fe(CN)_6$ is in a reduced state (Fig. S4) and the SOR active site is mostly in the oxidized state (Fig. S3). This suggests that a redox equilibrium has settled between the various oxidized and reduced species present at the end of the reaction, on a longer time scale than that observed by pulse radiolysis (Scheme 1). This most likely corresponds to an intra molecular electron transfer in the second intermediate, (Scheme 1), similarly to what occurs spontaneously when reduced SOR is mixed with ferricyanide, to form a SOR-$Fe^{3+}$-CN-$Fe^{2+}$(CN)$_5$ species (4, 11, 12). This complex is stable in solution, according to its $K_d$ value of 0.48±0.05 μM (Fig. S1). We note that the presence of only 0.7 mol of oxidized SOR generated per mol of $O_2^{\bullet-}$ together with the oxidation of 0.3 mol of free ferrocyanide per mol of $O_2^{\bullet-}$ into free ferricyanide indicates that the mixed-valence SOR-$Fe^{3+}$-CN-$Fe^{2+}$(CN)$_5$ has been partially reduced by free ferrocyanide (Scheme 1). From the data of Table 1, an equilibrium constant value of K≈0.1 can be calculated for this reduction process.

No $H_2O_2$ is produced at the end of the reaction. The absence of $H_2O_2$ does not result from any catalase activity of the reacting solution, nor from inhibition of the horseradish peroxidase enzyme used to quantify $H_2O_2$. Also, we observed no evidence for Fenton-like reactions, which would lead to $HO^\bullet$ production and destruction of the SOR-ferrocyanide complex. This is in line with the electronic balance of the SOR ferrocyanide solution at the end of the reaction, which indicates that $O_2^{\bullet-}$ has been reduced with only one electron. To form $HO^\bullet$ from $O_2^{\bullet-}$, two electrons are required.

Consequently, assuming a one-electron reduction process of $O_2^{\bullet-}$, a peroxide species should be formed, at least transiently, at the level of the first reaction intermediate. Reaction of this transient peroxide species with an organic solute present in the solution, formate or Tris buffer, could explain the absence of $H_2O_2$ formation at the end of the reaction. This is supported by the dependence of the rate constant $k_2$ on formate concentration (Fig S2). These data suggest that the peroxide intermediate species reacts with formate to form a performic acid at the end of the reaction, instead of $H_2O_2$ (Scheme 1). Because in SOR alone, the peroxide intermediate species formed during the catalytic cycle is quantitatively transformed into $H_2O_2$ and does not react with organic solutes, this reaction with formate appears to be specific to the SOR-ferrocyanide complex.

In a cellular context, such oxidation of an organic component to form an alkyl peroxide, weakly reactive as compared to the toxic $H_2O_2$ (1, 2), may result in a more efficient antioxidant property for the SOR-ferrocyanide complex than that of SOR alone. In fact, this is exactly what we observed *in vivo* on SOD deficient *E. coli* cells complemented with the *sor* gene. The *E. coli* mutant strain *sodA* *sodB recA* does not grow at all in the presence of oxygen unless complemented with *sor* or *sod* genes (14). With *sor* expression, the addition of ferrocyanide in the culture medium increased by a factor of three the aerobic survival of the *E. coli* mutant strain. This effect was specific to the presence of SOR and was not observed with SOD expression. We showed that in these conditions, the SOR-ferrocyanide complex is formed within the cells and that it can act as a catalyst to detoxify $O_2^{\bullet-}$. The soluble cell extracts contain NAD(P)H dependent reductases that can regenerate its reduced active form. Altogether, these data support the hypothesis that the SOR-ferrocyanide complex is more efficient in $O_2^{\bullet-}$ detoxification than SOR alone. The oxidized cellular components produced by the reaction of the SOR-ferrocyanide complex with $O_2^{\bullet-}$ could be much less damaging species than $H_2O_2$, possibly because they cannot be involved in the formation of the hydroxyl radicals by the Fenton reaction.

Addition of ferrocyanide as a cofactor to SOR provides a new detoxification mechanism for superoxide that may have a biological relevance. Although synthesis of ferrocyanide in cells has not been investigated to date, the presence of cyanide as a metal ligand has been well characterized in hydrogenases (15). Interestingly, hydrogenases are usually found in the same type of bacteria where SORs are naturally expressed and these cells possess enzymatic systems that can synthesize $CN^-$, HypE and HypF (16), and incorporate it into proteins. This leaves open a possible formation of ferrocyanide in these bacteria. One may anticipate that formation of the SOR-ferrocyanide complex could be associated with specific oxidative stress conditions, iron overload for example, which make cells highly and specifically sensitive to $H_2O_2$.

## MATERIALS AND METHODS

*Bacterial strains and plasmids.* *E. coli* strain QC2375 was previously described in (14). Plasmid pMJ25 is a pJF119EH derivative, in which the *sor* gene from *D. baarsii* is under the control of a tac promotor (14). The plasmid pCW-SODA (C. Weill, unpublished results) is a pMJ25 derivative, in which the *sor* gene is replaced by the *E. coli sodA* structural gene, allowing expression of the Mn SOD under the control of a tac promotor.

*Protein preparation*. The recombinant SOR from *D. baarsii* was purified as described in (4). The protein was isolated with a reduced active site, not oxidized in the presence of oxygen.

*Pulse radiolysis experiments*. Free radicals were generated by a 200 ns pulse of 4.5 MeV from a linear electron accelerator located at the Curie Institute, Orsay, France. $O_2^{\bullet-}$ was generated by scavenging radiolytically generated $HO^{\bullet}$ free radicals by 100 mM formate, in $O_2$ saturated solution, 10 mM Tris/HCl pH 7.6 (9). The doses per pulse were ca. 5 Gy ($[O_2^{\bullet-}]$=3 µM), calibrated with a thiocyanate dosimeter (9). The reaction was followed spectrophotometrically at 20 °C, in a 2 cm path length cuvette.

*γ-Ray radiolysis experiments*. γ-ray irradiations were carried out at 20 °C using a cobalt-60 source, at a dose rate of 17.6 Gy.min$^{-1}$. The SOR solutions were made up in 10 mM Tris/HCl buffer pH 7.6, 100 mM sodium formate and saturated with pure $O_2$ at room temperature. Concentrated potassium hexacyanoferrate (II) was added to the solution just prior to irradiation. The duration of irradiation was 5 min (88 Gy). Immediately after irradiation, UV-visible spectra of the solution were recorded in parallel with that of a non-irradiated solution kept under the same conditions. Hydrogen peroxide production was determined immediately after irradiation using the leuco crystal violet horseradish peroxidase method as described in (17).

*Electrospray ionization mass spectroscopy*. Mass spectra were obtained on a Perkin-Elmer Sciex API III+ triple quadrupole mass spectrometer equipped with a nebulizer-assisted electrospray source operating at atmospheric pressure. Samples were in 10 mM ammonium acetate.

*FTIR spectroscopy*. FTIR absorption spectra were recorded at 4 cm$^{-1}$ resolution, using a Bruker 66 SX spectrometer equipped with a KBr beam splitter and a nitrogen-cooled MCT-A detector. Each spectrum corresponds to 300 scans. 15 µL of the liquid sample was deposited between two $CaF_2$ windows separated with a ≈25 µm spacer.

ACKNOWLEDGEMENTS

This work has been partly supported by the Toxicologie Nucléaire program from the CEA. F.P.M.-H. acknowledges a postdoctoral fellowship from the CEA. We acknowledge J. Vicente for irradiation experiments, D. Lemaire for mass spectroscopy, D. Bourgeois for helpful discussions and M. Fontecave for constant support on this work.

FIGURE LEGENDS

**Fig. 1.** Crystal structure of the active site of the SOR from *D. baarsii* in complex with ferrocyanide (8). The coordination of $Fe(CN)_6$ to the SOR iron site occurs through a cyano-bridge. The cyanide moieties are and stabilized by hydrogen bonds (red dotted line) and van der Walls interactions (thin black dotted lines).

**Fig. 2. A**. First 300 µs of the reaction of SOR (100 µM) with $O_2^{\bullet-}$ (3 µM), generated by pulse radiolysis, followed at 580 nm, in presence of 0, 1 and 1.2 to 10 molar equivalents of ferrocyanide with respect to SOR, at pH 7.6. The lines were calculated for best fit to an exponential model. **B**. Dependence of the observed rate constants $k_{1app}$ on ferrocyanide concentration. The inset shows the dependence of $k_{1app}$ versus SOR concentration, in the presence of 5 molar equivalents of ferrocyanide with respect to SOR.

**Fig. 3.** First 19 ms of the reaction of SOR (100 µM) with $O_2^{\bullet-}$ (3 µM), generated by pulse radiolysis, followed at 580 nm, in presence of 0, 1 and 1.2 to 10 molar equivalents of ferrocyanide with respect to SOR, at pH 7.6. The lines were calculated for best fit to a bi-exponential model. The inset shows the dependence of $k_{2app}$ versus SOR concentration, in the presence of 5 molar equivalents of ferrocyanide with respect to SOR.

**Fig. 4.** Transient absorption spectra (2 cm path-length cuvette) formed upon reaction of SOR (100 µM) in the presence of 0 and 5 molar equivalents of ferrocyanide with $O_2^{\bullet-}$ (3 µM), generated by pulse radiolysis, at pH 7.6. **A**. Reconstituted spectra of the first reaction intermediates; (□) in the absence of ferrocyanide 50 µs after the pulse; (●) in the presence of 5 molar equivalents of ferrocyanide 300 µs after the pulse. **B**. Reconstituted spectra of the second reaction intermediates; (□)

in the absence of ferrocyanide 10 ms after the pulse; (●) in the presence of 5 molar equivalents of ferrocyanide 10 ms after the pulse. Dotted line is the spectrum of a ferricyanide species at 3 μM.

**Fig. 5.** FTIR difference spectra calculated from the absorption spectra of SOR-ferrocyanide solutions recorded after *minus* prior to the reaction with 55 μM $O_2^{\bullet-}$, generated by γ-ray radiolysis. The solution contained 100 μM of SOR and various concentration of ferrocyanide, at pH 7.6. Upper spectrum: sample with 2 molar equivalents of ferrocyanide with respect to SOR. Lower spectrum: sample with 5 molar equivalents of ferrocyanide with respect to SOR. The quantification of the ferrocyanide (2037 cm$^{-1}$) and ferricyanide (2115 cm$^{-1}$) bands were deduced from FTIR spectra recorded in the same conditions with solutions of known concentration for each species.

Scheme 1. Proposed mechanism of the SOR-ferrocyanide complex with $O_2^{\bullet-}$.

Table 1. Products for the reaction of the SOR-ferrocyanide complex with 55 µM of $O_2^{\bullet-}$, generated by γ-ray radiolysis (88 Gy at 17.6 Gy/min), at pH 7.6. The SOR solution (100 µM) contained various molar equivalent of ferrocyanide compared to the SOR iron active site. Each value represents the mean of at least three independent experiments.

| Ferrocyanide (molar eq) | $[H_2O_2]_{final}$[a] (µM) | $[H_2O_2]_{corr}$[b] (µM) | [SOR] oxidized[c] (µM) | [Ferricyanide][c] (µM) |
|---|---|---|---|---|
| 0 | 62±4 | 56 | 57 | 0 |
| 0.5 | 44±4 | 38 | nd | nd |
| 1 | 30±1 | 24 | 50 | 14 |
| 2 | 6.1±0.5 | 0 | 47 | 15 |
| 3 | 4.7±1.2 | 0 | 44 | 17 |
| 5 | 7.2±0.5 | 1 | 42 | 18 |
| 10 | 4.7±2.0 | 0 | nd | nd |
| SOR+5eq ferrocyanide +50 µM $H_2O_2$ | 54±1 | 48 | nd | nd |
| No SOR, ferrocyanide 200µM | 34±2 | 28 | / | 0 |

[a] dosage of $H_2O_2$ with the leuco crystal violet method. [b] corrected from the radiolytic yield for $H_2O_2$ = 6.1 µM. [c] determined from the spectra of Fig. S3, using $\varepsilon_{644\,nm}$ = 1900 $M^{-1}$ $cm^{-1}$ for the oxidized SOR active site and $\varepsilon_{420nm}$ = 1010 $M^{-1}$ $cm^{-1}$ for the ferricyanide.

Table 2. Effect of ferrocyanide and SOR production on the aerobic survival of a *sodA sodB recA E. coli* mutant (QC 2375). Anaerobic culture of QC 2375 transformed with pJF119EH, pMJ25 or pCWSOD were plated on LB medium plus $2 \cdot 10^{6}$ M IPTG and different concentrations of ferrocyanide, under anaerobic and aerobic conditions. Colonies were counted after 24 h incubation at 37 °C.

| [Ferrocyanide] (mM) | Aerobic survival (%)[a] | | |
|---|---|---|---|
| | QC 2375 pJF119EH (control vector) | QC 2375 pMJ25 (Sor$^+$) | QC 2375 pCWSOD (Sod$^+$) |
| 0 | 0.003 | 21±4[b] | 23±6[c] |
| 1 | 0.003 | 72±7[c] | 34±12[c] |
| 5 | nd | 51±10[b] | 17±5[b] |

[a] Survival was calculated as the ratio of the number of colonies under aerobic conditions to those under anaerobic conditions. Values are the means of four experiments. 100% corresponds to $1.8\text{-}2.3 \cdot 10^{8}$ colonies. [b] Tiny colonies, became larger after 48 h. [c] Large colonies after 24 h.

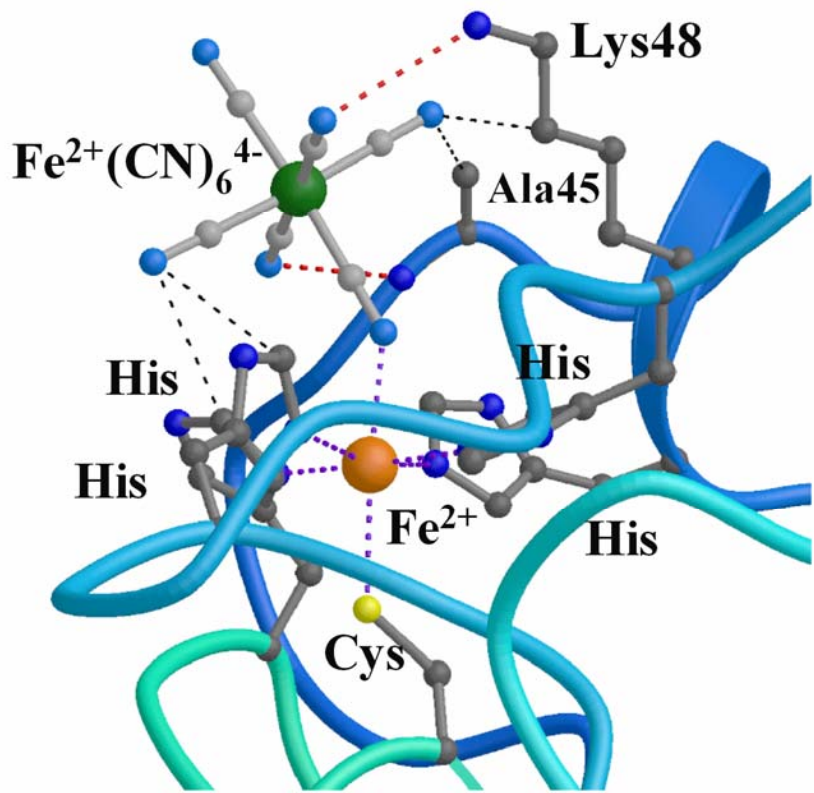

Fig. 1

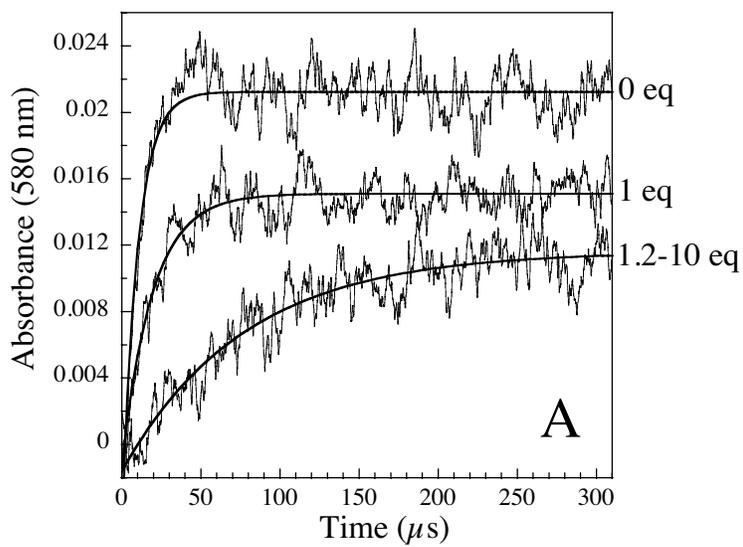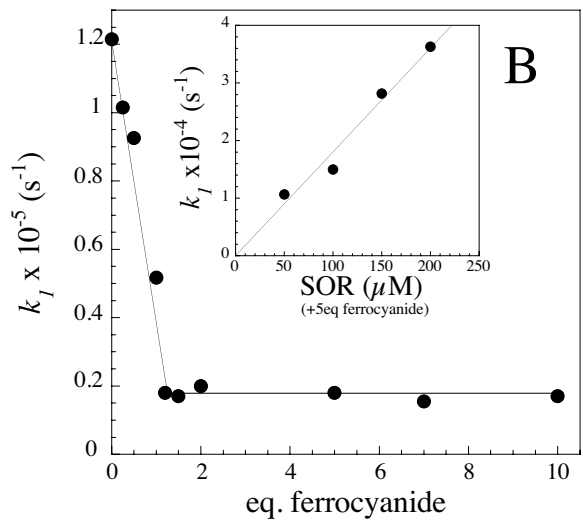

Fig. 2

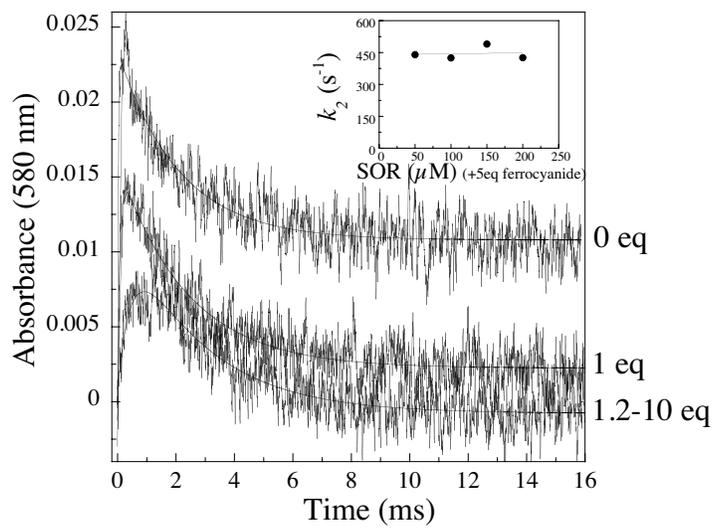

Fig. 3

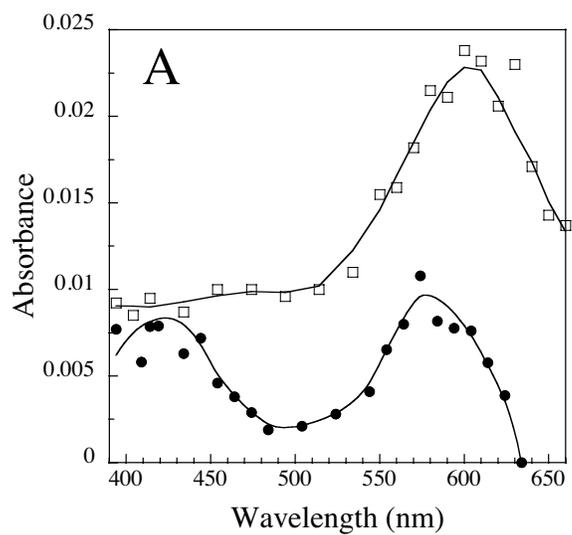 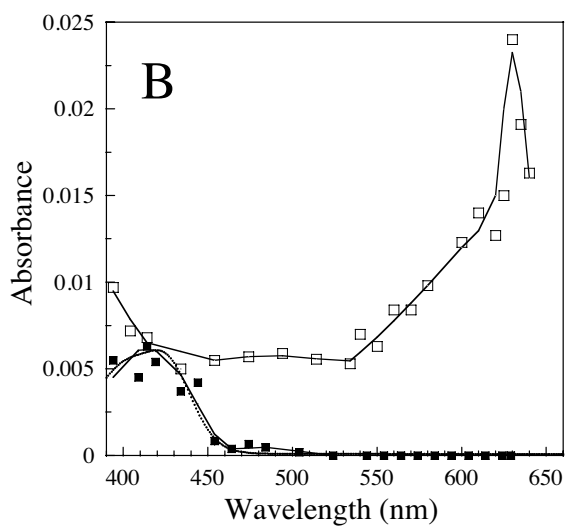

Fig. 4

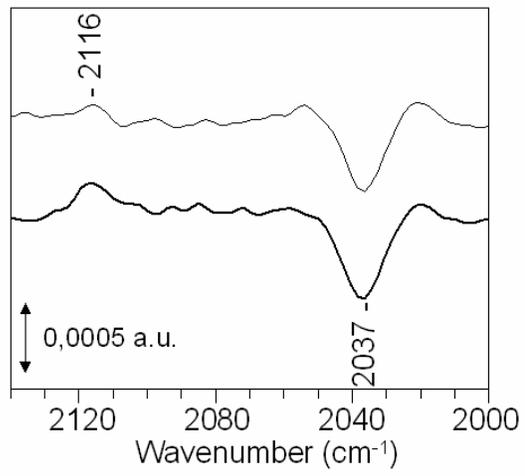

Fig. 5

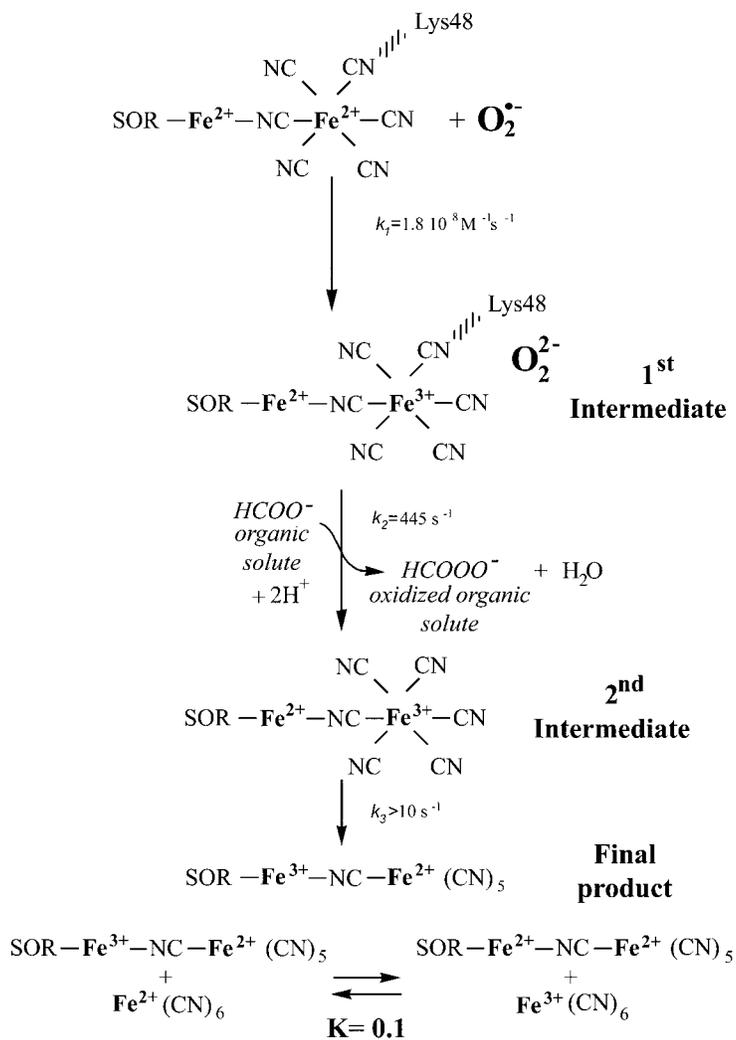

Scheme 1